\documentclass[preprint,showpacs,preprintnumbers,amsmath,amssymb]{revtex4}
\usepackage{amssymb}
\usepackage{graphicx}
\usepackage{dcolumn}
\usepackage{bm}
\begin{document}
\title{Interacting bosons in an optical lattice: Bose-Einstein
    condensates and Mott insulator}
\author{O. Fialko}
\author{Ch. Moseley}
\author{K. Ziegler}
\altaffiliation{Institut f\"ur Physik, Universit\"at Augsburg, Germany}
\begin{abstract}
A dense Bose gas with hard-core interaction is considered in 
an optical lattice. We study the phase diagram in terms of a 
special mean-field theory that describes a Bose-Einstein condensate
and a Mott insulator with a single particle per lattice site
for zero as well as for non-zero temperatures.
We calculate the densities, the excitation spectrum and the
static structure factor for each of these phases. 
\end{abstract}

\pacs{05.30.Jp, 03.75.Hh, 03.75.Lm}

\maketitle

\section{Introduction}

Optical lattices have opened an exciting field of physics. We are
expecting new phenomena in comparison with continuous
systems due to the lack of Galilean invariance. Among the most
obvious consequences of the  lattice structure is the formation of 
lattice-commensurate ground states like the 
Mott insulating phase. The latter was observed in experiments
\cite{exper1,exper2}.

\begin{figure}[h]
\begin{center}
\scalebox{1.4}{\includegraphics[width=12cm]{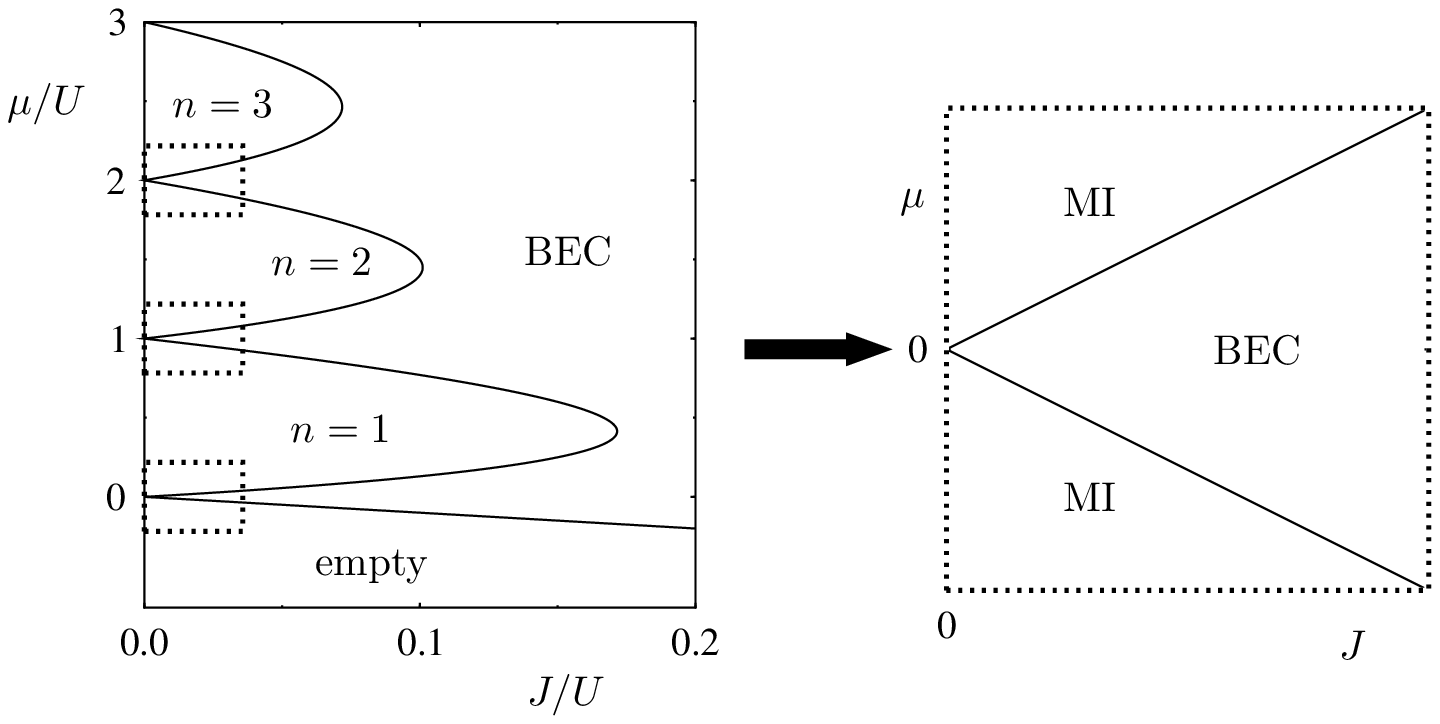}}
\end{center}
\caption{A projection of the phase diagram of the Bose-Hubbard model in the vicinity of the point, where the 
two Mott lobes meet. $\mu$ and $J$ are in arbitrary energy units after the projection.}
\label{projection}
\end{figure}

Light scattering on a Bose gas is strongly affected by the
nature of the quasiparticles \cite{Gorlitz,ketterle1}. This provides a
useful experimental tool to distinguish between different phases of the 
Bose gas. We expect, for instance, that light scattering by the 
gapless quasiparticle spectrum of the Bose-Einstein condensate 
is quite different from light scattering in the gapful 
quasiparticle spectrum of the Mott insulator (MI). A physical quantity
that is directly related to light scattering is the structure factor
\cite{ketterle1,Pitaevskii,brand,Ates,huber,batrouni} which was measured 
in the case of a BEC \cite{Gorlitz}. This is also of interest in the Mott insulating
phase \cite{ana1, ana2}. In this paper we will study
the static structure factor. Of particular interest is its behavior
near the transition from the BEC to the MI. We will also study the phases and phase 
transitions for 
zero temperature as well as for nonzero temperature. For this purpose we will consider here
a hard-core Bose gas in an optical lattice.  Our model can be understood
as a projection of the more general Bose-Hubbard model in the vicinity
of those points of the phase diagram, where two adjacent Mott lobes meet 
(Fig. \ref{projection}).
This is similar to the picture which was applied to the tips of the Mott lobes in a recent paper
by Huber et al. \cite{huber}. It is based on the following idea.
The number of bosons per site is fixed in the Mott state.
For adjacent Mott lobes this means that the corresponding Mott states differ exactly
by one boson per site. Now we consider two adjacent lobes with $n$ and $n+1$ ($n\ge0$ bosons per
site), respectively and assume that the chemical potential is fixed such that the ground
state is the Mott state with $n$ particles per site.
Low-energy excitations of this state for a grand-canonical system are 
states, where one or a few sites (let's say $k\ge1$ sites) have $n+1$ bosons, 
all other sites have $n$ bosons. The $k$ excessive bosons are relatively free to move from site
to site on top of the $n$ Mott state.
Therefore, the physics of these excitations can be described approximately by the tunneling
of the $k$ excessive bosons alone.
Due to the repulsion of order $U$, assumed to be not too small, it is unlikely that a 
site with $n+2$ bosons is created. Consequently, these excessive bosons form a hard-core
Bose gas.

A hard-core  boson can be represented by a
pair of locally coupled spin-1/2 fermions. Here we will use the model 
introduced to study the dissociation of bosonic molecules into pairs 
of spin-1/2 fermionic atoms \cite{Ziegler}. A functional integral
with a two-component complex field was derived for this model that 
allowed a mean-field approximation. The latter
revealed a zero-temperature phase diagram for a grand-canonical
enesemble of bosons with three phases: an empty phase, a BEC and a MI.
This result is remarkable, since previous mean-field
calculations, using the same type of fermionic model as starting point,
(e.g. using an $N\to\infty$ limit \cite{N-comp}) did not give
the entire phase diagram by ignoring the MI phase. Aparently,  
the form of the phase
diagram depends crucially on the type of Hubbard-Stratonovich transformation
that replaces the fermionic (i.e. Grassmann) fields by bosonic (i.e. complex)
fields.
 
The structure of this paper is as follows. In Sec. 2 we discuss briefly
the model and some related physical quantities. A mean-field approximation
is used in Sec. 3, where we derive the phase diagram. Gaussian fluctuations
around the mean-field solution and its consequences are
considered in Sec. 4. In this section we calculate the static
structure factor. And in Sect. 5 we discuss our results. Details of our 
calculations are given in the Appendices A and B.

\section{The Model}

A hard-core Bose gas is considered on a $d$-dimensional
hyper-cubic lattice with $\cal{N}$ sites. Bosons can tunnel
between these sites from a lattice site $r$ to any of 
the nearest-neighbor sites $r^{\prime}$. Each boson consists of two
fermions. Single fermions in this model can not exist (i.e., in 
contrast to the model studied in Ref. \cite{Ziegler} we do not 
allow dissociation of the bosonic molecules). The
Hamiltonian of the $d$-dimesional system is
\begin{equation}
\hat{H}=-\frac{J}{2d}\sum_{\langle
r,r^{\prime}\rangle}c_{r\uparrow}^{\dagger}c_{r^{\prime}\uparrow}
c_{r\downarrow}^{\dagger}c_{r^{\prime}\downarrow}-\mu\sum_{r}
\sum_{\sigma=\uparrow\downarrow}c_{r\sigma}^{\dagger}c_{r\sigma},
\label{model}
\end{equation}
where the sum in the first term on the right-hand side is
restricted to nearest neighbors, and $c^{\dagger}_{r\sigma}$ and
$c_{r\sigma}$ are the creation and annihilation operators of the
fermion with spin $\sigma$ at site $r$, respectively. The first
term describes tunnelling of local fermion pairs in the optical
lattice between nearest-neighbor sites with rate $J$. The chemical
potential $\mu$ controls the number of particles in a
grand-canonical ensemble. The latter is given by the partition
function
\begin{equation}
Z=\mbox{Tr} e^{-\beta\hat{H}}.
\end{equation}
The partition function can also be written in terms of a Grassmann
integral \cite{Negerle} as
\begin{equation}
Z=\int e^{-A(\psi,\bar{\psi})}D[\psi,\bar{\psi}] \label{ZModel}
\end{equation}
with imaginary time-dependent conjugated Grassmann fields $\psi$,
$\bar{\psi}$ and the action
\begin{equation}
A=\int_{0}^{\beta}d
t\left[\sum_{r}(\psi_{r}^{1}\partial_{t}\bar{\psi}_{r}^{1}+\psi_{r}^{2}
\partial_{t}\bar{\psi}_{r}^{2})-\mu\sum_{r}(\psi_{r}^{1}\bar{\psi}_{r}^{1}
+\psi_{r}^{2}\bar{\psi}_{r}^{2})-\frac{J}{2d}\sum_{\langle
r,r^{\prime}\rangle}\psi_{r}^{1}
\bar{\psi}_{r^{\prime}}^{1}\psi_{r}^{2}\bar{\psi}_{r^{\prime}}^{2}\right].
\label{SModel}
\end{equation}

Applying the concept of linear response to the quasiparticles, we
obtain the dynamic structure factor with the definition
\cite{Pitaevskii}
\begin{equation}
S(q,\omega)=\sum_{n,m}e^{-\beta\hbar\omega_{m}}|\langle n
|\hat{\rho}_{q}^{\dagger}-
\langle\hat{\rho}_{q}^{\dagger}\rangle|m  \rangle|^{2}\delta(\hbar
\omega - \hbar\omega_{n,m})
\label{def_of_S}
\end{equation} 
where $\rho^{\dagger}_{q}$ is the Fourier transform of 
the atomic density operator at 
wave vector $q$ and $\omega_{n,m}$ is the frequency difference between energy
level $m$ and $n$. We treat an excited state $n$ as one of the
quasiparticles energy level with $\hbar\omega_{n,0}=\epsilon_{n}$
within our approximation. Then the static structure factor is defined as \cite{Pitaevskii}
\begin{equation}
S(q)=\frac{\hbar}{N} \int S(q,\omega)d\omega=\frac{1}{N}(\langle \rho_q \rho_{-q} 
\rangle-|\langle \rho_q \rangle|^2).
\label{sq}
\end{equation}
For small but nonzero $T$, when quasiparticles can be treated
as noninteracting  within the accuracy of the Bogoliubov approach for the 
weakly interacting Bose gas modified for an optical lattice we get
\cite{Pitaevskii}
\begin{equation}
S(q)=\frac{Jg_{q}}{\epsilon_{q}}\coth\frac{\beta\epsilon_{q}}{2},
\label{Struct}
\end{equation}
where 
\begin{equation}
g_{q}=1-\frac{1}{d}\sum_{i=1}^{d}\cos q_{i}.
\label{g}
\end{equation}
For $T=0$
\begin{equation}
S(q)=\frac{Jg_{q}}{\epsilon_{q}},
\end{equation}
where $\epsilon_{q}$ is the quasiparticle spectrum.

This result was obtained within the mean-field approximation to the weakly 
interacting Bose gas. It is interesting to compare
this result with the result obtained by a direct calculation for our model within our 
mean-field approach. Using Eq.(\ref{sq}) the static structure factor can be calculated as 
\cite{Feynman}
\begin{equation}
S(q)=\frac{1}{N}\sum_{r,r^{\prime}}C_{r,r^{\prime}}e^{iq(r-r^{\prime})},
\end{equation}
where the truncated density-density correlation function is
\begin{equation}
C_{r r^{\prime}}=\langle n_{r}n_{r^{\prime}}\rangle-\langle
n_{r}\rangle\langle n_{r^{\prime}}\rangle. 
\end{equation}
Introducing spatial coordinates $p=\{r,t\}$ the static structure factor reads
\begin{equation}
S(q)\equiv\frac{1}{\beta N}
\sum_{r,r^{\prime}}\int_{0}^{\beta}d t \int_0^\beta dt^{\prime}  C_{rt,r^{\prime}t^{\prime}}e^{iq(r-r^{\prime})}. \label{Sq}
\end{equation} 
This expression will be studied within mean-field theory and Gaussian fluctuations.

\section{Mean-Field Approximation: Phase Diagram}

The mean-field approximation cannot be directly applied to the Grassmann fields.
Therefore, we perform a
Hubbard-Stratonovich transformation in order to replace the Grassmann fields
by  complex fields. This leads to an effective
action depending on complex fields. As we already mentioned in the
Introduction, the form of the Hubbard-Stratonovich transformation
is not unique. Here we use the version given in Ref. \cite{Ziegler}
because it provides the proper phase diagram.
First of all we decouple the fourth order term in the action using 
the identity
\[
\exp\left( \frac{J}{2d}\sum_{\langle r,r^{\prime}\rangle}\psi_{r}^{1}
\bar{\psi}_{r^{\prime}}^{1}\psi_{r}^{2}\bar{\psi}_{r^{\prime}}^{2}\right) 
=\int[d\phi][d\chi]\exp\left[-\sum_{r,r^{\prime}}
\bar{\phi}_{r}\hat{v}^{-1}_{r,r^{\prime}}\phi_{r^{\prime}}
-\frac{1}{2J}\sum_{r}\bar{\chi}_{r}\chi_{r}-\right.
\]
\begin{equation}
\left.-\sum_{r}(i\phi_{r}+\chi_{r})\psi_{r,1}\psi_{r,2}- 
\sum_{r}(i\bar{\phi}_{r}+\bar{\chi}_{r})\bar{\psi}_{r,1}\bar{\psi}_{r,2}\right].
\label{decoupling}
\end{equation}
Summation over $t$ in the last formula is implicitly assumed. In this formula the
expression
\begin{equation}
\hat{v}_{r,r^{\prime}}=J\left(\frac{1}{2d}\delta_{|r-r^{\prime}|,1}+2\delta_{r,r^{\prime}}\right)
\end{equation}
was introduced. Moreover, the introduction of the auxiliary field $\chi$ makes the matrix
$\hat{v}$ positive definite.

A subsequent integration over Grassmann fields leads to
\begin{equation}
A_{\rm eff}=\int_{0}^{\beta}d t\left\{\sum_{r,r^{\prime}}
\bar{\phi}_{r}\hat{v}^{-1}_{r,r^{\prime}}\phi_{r^{\prime}}
+\frac{1}{2J}\sum_{r}\bar{\chi}_{r}\chi_{r}-\log\det \hat{\bf{G}}^{-1}\right\}
\end{equation}
with
\[
\hat{\bf{G}}^{-1}=\left(\begin{array}{cc} -i\phi-\chi &\partial_{\tau}+\mu  \\
\partial_{\tau}-\mu & i\bar{\phi}+\bar{\chi}  \end{array} \right),
\ \ \hat{v}=J(\hat{\omega}+2\hat{1}), \ \
\hat{\omega}_{r,r^{\prime}}=\left\{ \begin{array}{ll}1/2d, &
r,r^{\prime}-\mbox{nearest neighbors} \\ 0, &
\mbox{otherwise}\end{array}\right. .
\]
The partition function now reads
\begin{equation}
Z=\int e^{-A_{\rm eff}(\phi,\chi)}D[\phi,\chi] .
\end{equation}
We can perform a saddle-point integration to calculate
physical quantities. From a physical point of view we have to
minimize our action to get the classical trajectory 
(i.e. the macroscopic wavefunction) of our system.
Fluctuations around this trajectory are caused by thermal and
quantum effects.
The mean-field solution characterizes the condensed phase, in which
$|\phi|$ has a nonzero value. Fluctuations describe quasiparticle
excitations above the condensate.
In order to proceed within a mean-field approximation we assume that 
quantum fluctuations are small.

Minimization of the action gives us two coupled linear equations
between the complex fields $\phi$ and $\chi$:
\begin{equation}
\delta A_{\rm eff}=0 \ \ \Rightarrow \ \
\left\{\begin{array}{l}\phi=3JG(\phi-i\chi) \\
\chi=-i2JG(\phi-i\chi) \end{array}  \right. ,
\end{equation}
where
\[
G=\frac{1}{\beta}\sum_{\omega_n}\frac{1}{|\phi|^{2}/9+\mu^{2}/4+\omega_{n}^{2}}.
\]
A solution of these equations is
\begin{equation}
G=\frac{1}{J}, \ \ \chi=-\frac{2i\phi}{3}.
\end{equation}
In these equations $\omega_{n}=(2n+1)\pi/\beta$ are the Matsubara
frequencies of the fermions. This leads to (we perform
the rescaling $4|\phi|^{2}/9\rightarrow |\phi|^{2}$)
\begin{equation}
J=\sqrt{\mu^{2}+|\phi|^2}
\left(\frac{e^{\beta\sqrt{\mu^{2}+|\phi|^{2}}/2}+1}{e^{\beta
\sqrt{\mu^{2}+|\phi|^{2}}/2}-1} \right)
\label{J}
\end{equation}
and within the saddle-point integration we obtain the expression for the local density
\begin{equation}
n=\frac{1}{\beta\cal{N}}\frac{\partial \log Z}{\partial \mu}
=\frac{1}{2}+\frac{1}{2}\frac{\mu}{\sqrt{\mu^{2}+|\phi|^{2}}}
\left(\frac{e^{\beta\sqrt{\mu^{2}+|\phi|^{2}}/2}-1}
{e^{\beta\sqrt{\mu^{2}+|\phi|^{2}}/2}+1} \right).
\label{n}
\end{equation}
It should be noticed that this expression for $J=0$ and $T=0$ gives correct 
result, which can alone be obtained by direct calculations. The corresponding phase 
diagram is depicted in Figs. 1
and 2 with the phase boundary given by $\phi=0$.

With the result given in Eq. (\ref{n}) we can evaluate the local
compressibility:
\begin{equation}
k_{r}=\left(\frac{\partial n_{r}}{\partial \mu}\right)_{T}.
\end{equation}
For $T=0$ and small $|\phi|^2$ it reads
\begin{equation}
k_{r}\approx\frac{|\phi|^2}{2\mu^3}.
\label{compres}
\end{equation}
The local compressibility is not divergent within our mean-field
approximation, which is typical in the case of optical lattices
\cite{Jaksch, Batr, Rigol}.

We see that this model gives three phases. It can describe
both the dilute regime and the dense regime, as a consequence we
have the empty phase along with the MI phase. 

A few words about the type of the phase transitions. Near the
phase transition the order parameter is small, so that we may apply a 
perturbative expansion :
\begin{equation}
A_{\rm eff}=\underbrace{\frac{1}{J}\frac{|\phi|^{2}}{9}-
\frac{1}{\beta}\sum_{\omega_n}\frac{1}{\mu^{2}+\omega_n^{2}}\frac{|\phi|^{2}}{9}}_{\mbox{can be 
negative}}
+\underbrace{\frac{1}{2\beta}
\sum_{\omega_n}\frac{1}{(\mu^{2}+\omega_n^{2})^{2}}\frac{|\phi|^{4}}{81}}_{>0}+O(|\phi|^{6})
\label{Landau}
\end{equation}
The first term can be negative, such that the order parameter may
differ from zero. The order parameter vanishes continueously 
indicating that we have a second order phase transition.

\section{Gaussian Fluctuations around Mean-Field Solution}

The complex fields $\phi_{x}$ and $\chi_{x}$ are expected to
fluctuate about the SP solution due to thermal and quantum effects. Denote
$\Delta=i\phi+\chi$ and $\bar{\Delta}=i\bar{\phi}+\bar{\chi}$,
then
\begin{equation}
\hat{\bf{G}}^{-1}=\hat{\bf{G}}^{-1}_{0}+\left(\begin{array}{cc}-\delta\Delta
& 0  \\ 0 & \delta\bar{\Delta} \end{array} \right),
\label{ExpansionGreen}
\end{equation}
where
\[
\hat{\bf{G}}^{-1}_{0}=\left(\begin{array}{cc} -\Delta_{0}
&\partial_{\tau}+\mu  \\ \partial_{\tau}-\mu & \bar{\Delta}_{0}
\end{array} \right).
\]
Applying the Taylor expansion $\ln(1+x)=x-x^{2}/2+...$ we get
\[
\log \det \hat{\bf{G}}^{-1}=\mbox{tr} \ln
\hat{\bf{G}}^{-1}=\mbox{tr} \ln
\left[\hat{\bf{G}}_{0}^{-1}+\left(\begin{array}{cc}-\delta\Delta
& 0  \\ 0 & \delta\bar{\Delta} \end{array} \right)\right]\approx
\]
\begin{equation}
\approx\mbox{tr} \ln
\hat{\bf{G}}^{-1}_{0}-\frac{1}{2}\mbox{tr}\left[\hat{\bf{G}}_{0}
\left(\begin{array}{cc}
-\delta\Delta & 0 \\ 0 & \delta\bar{\Delta} \end{array}
\right)\right]^{2} .
\label{ExpansionLog}
\end{equation}
Calculating the trace in $p=\{q,\omega\}$ representation we get
\begin{equation}
Z\sim\int D[\delta\phi]\exp\left[-\delta A_{\rm eff} \right],
\end{equation}
where  $\delta A_{\rm eff}$ is given in Appendix A.
This has the form of an inverse Green's function $\hat{G}^{-1}$ with
imaginary time for quasiparticles, which is described by fields
$\delta\phi$. By applying the spectral representation of the Green's
function we can identify the poles of function $\hat{G}$, namely
$i\omega$ with the excitation spectrum of the quasiparticles.
The exact solution in the condensed phase is
\begin{equation}
\epsilon_{q}=\sqrt{g_{q}(J^{2}-\mu^{2})+g_{q}^{2}\mu^{2}}.
\label{spectrum}
\end{equation}
In the dilute regime, where $\mu+J$ is small, and for small $q$ we
get the Bogolubov spectrum with an effective mass of the bosons
$\sim 1/J$ and sound velocity $\approx\sqrt{2J\tilde{\mu}}$ (we
used $\tilde{\mu}=J+\mu$). This has the same form as the
quasiparticle spectrum obtained in the \cite{N-comp}.

In the empty phase and in the MI phase we have
excitations with a gap:
\begin{equation}
\epsilon_{q}=|\mu|-J+Jg_{q} .
\label{M}
\end{equation}
The symmetric form of the results for the MI phase and
for the empty phase is due to the particle-hole symmetry of our
model.

In the dilute regime (small $\tilde{\mu}=J+\mu$) and for low 
temperatures, including fluctuations, we obtain
\begin{equation}
n_{0}=n-\sum_{q\ne 0}n_{q}\approx \tilde{\mu}\int\frac{dq}{J
g_{q}}-(k_B T)^{\frac{3}{2}}\int\frac{d^{3}q}{(2\pi)^{3}}\frac{1}{e^{
g_{q}}-1}
\end{equation}
This result is in good agreement with a weakly interacting Bose
gas with an effective chemical potential $\tilde{\mu}$
\cite{Popov}.

\subsection{Quantum Fluctuations versus Thermal Fluctuations}
By defining the condensed density in the following way
\[
n_o=\lim_{|r-r^{\prime}|\rightarrow\infty}\langle c^{\dagger}_{r\uparrow}
c_{r^{\prime}\uparrow}c^{\dagger}_{r\downarrow}c_{r^{\prime}\downarrow}  \rangle = 
\lim_{|r-r^{\prime}|\rightarrow \infty}\langle \psi_r^1  \bar{\psi}_{r^{\prime}}^1
\psi_r^2\bar{\psi}_{r^{\prime}}^2
\rangle
\]
we arrive at the expression for the condensed density, which can be calculated as
\[
\lim_{|r-r^{\prime}|\rightarrow\infty}\int_0^{\beta} d t \int_0^\beta d t^{\prime}
\left\langle\frac{-\Delta_{rt}\bar{\Delta}_{r^{\prime}t^{\prime}}}
{(-\partial_t^2+\mu^2/4-\Delta_{rt}\bar{\Delta}_{rt})(-\partial_t^{\prime
2}+\mu^2/4-\Delta_{r^{\prime}t^{\prime}}\bar{\Delta}_{r^{\prime}t^{\prime}})}\right\rangle .
\]
Then fluctuating fields $\Delta=\Delta_0+\delta\Delta$, expanding the above expression up to the 
second 
order of $\delta\Delta$ and using the Green's function given in Appendix 
A,  we arrive at \begin{equation}
n_{o}=\frac{(J^2-\mu^2)}{4 J^2}+\delta n_o,
\label{no_quantum}
\end{equation}
where the correction due to the quantum fluctuations to the mean-field result is
\[
\delta n_{o}=-\frac{(J^2-\mu^2)\mu^2}{J^3}\int \frac{d^d k}{(2\pi)^d} \frac{B_k^2
g_k}{\epsilon_k}+\frac{(J^2-\mu^2)}{4 J^3}\int \frac{d^d k}{(2\pi)^d} B_k 
\epsilon_k-\frac{(J^2-\mu^2)^2}{4 J^3}\int 
\frac{d^d k}{(2\pi)^d}
\frac{B_k^2}{\epsilon_k}+
\]
\[
+\frac{3
(J^2-\mu^2)^2\mu^2}{4 J^5}\int \frac{d^d k}{(2\pi)^d} \frac{B_k^2 g_k}{\epsilon_k}
\]
The main correction due to the thermal fluctuations are already included in our mean-field theory, 
where the condensed density is given by
\begin{equation}
n_o=\frac{|\phi|^2/4}{J^2},
\label{no_thermal}
\end{equation}
and $|\phi|^2$ can be determined from Eq. (\ref{J}).

The effect of quantum fluctuations and thermal fluctuations is depicted in Fig. \ref{fluct}.  We see 
that both of them lead to a depletion of the 
condensate, but the quantum depletion alone does not change the transition points.

\subsection{Static Structure Factor} 

The static structure factor is a Fourier transform of the truncated density-density correlation
function. The latter contains nonlocal term
\begin{equation}
\langle 
\psi_{r}^{1}\psi_{r}^{2}\bar{\psi}_{r}^{1}\bar{\psi}_{r}^{2} 
\psi_{r^{\prime}}^{1}\psi_{r^{\prime}}^{2}\bar{\psi}_{r^{\prime}}^{1}\bar{\psi}_{r^{\prime}}^{2} 
\rangle=\int_0^\beta dt \int_0^\beta dt^{\prime} \left\langle 
\frac{1}{(-\Delta_{rt}\bar{\Delta}_{rt}-\partial_{t}^{2}+\mu^2/4)(-\Delta_{r^{\prime}
t^{\prime}}\bar{\Delta}_{r^{\prime}t^{\prime}}-\partial_{t^{\prime}}^{2}+\mu^2/4)}   
\right\rangle
\label{correl}
\end{equation}
In mean-field approximation the truncated density-density correlation function should vanish 
\cite{Ziegler}. Taking into account fluctuations, we write again $\Delta=\Delta_{0}+\delta\Delta$, 
 $\bar{\Delta}=\bar{\Delta}_{0}+\delta\bar{\Delta}$ and substitute it into Eq.(\ref{correl}). Then 
expanding up to the second order and Fourier 
transforming the obtained expression, after a direct
but lengthy calculations we get for small wave vectors $q$ and for low 
temperatures $T$ in the BEC 
phase
\begin{equation}
S(q)\sim\frac{(J^2-\mu^2)}{J^2n}\frac{Jg_{q}}{\epsilon_{q}}\coth\frac{\beta\epsilon_{q}}{2},
\end{equation} 
where $n$ is a total density of particles.

In the dilute regime, i.e. close to the empty phase, when $n\sim (J+\mu)/J$ and $J-\mu\approx 2J$ 
we obtain
\[
S(q)\sim \frac{Jg_{q}}{\epsilon_{q}}\coth\frac{\beta\epsilon_{q}}{2},
\]
which is in agreement with the well-known result for the weakly interacting Bose gas 
\cite{Pitaevskii}. In the 
dense regime, i.e. close to the Mott phase when $n\approx 1$, the static structure factor 
vanishes. The dependence of the static structure factor on $q$ in the dilute regime is depicted
in Fig. \ref{Sdilute}. 

The static structure factor measures the density-density
correlations in $q$ space. Assuming that the thermal energy is much
larger than the gap (i.e. $k_{B}T\gg \Delta$), the
maximum of the structure factor appears at $q$  values for which the following 
condition holds
\begin{equation}
Jg_{q}\approx \Delta, 
\label{maximum}
\end{equation}
where $\Delta=\mu-J$ is an energy gap in the excitation spectrum. Thus fluctuations begin to feel each 
other
at the distance equals to their effective size $r\sim 1/q\sim
1/\sqrt{\mu-J}$.

Spatial correlations can be calculated by Fourier transforming the
static structure factor. Near the phase transition for $T=0$ 
 we have in the condensed phase for large $r$(see Appendix B)
\begin{equation}
C_{r,0}=\sum_{q}S(q)e^{iqr}\sim\frac{1}{r^{d+1}}.
\end{equation}
A similar behavior was found for a one-dimensional lattice in
\cite{Ates}. 

\section{Discussion}
A paired-fermion model for bosons with a hard-core interaction was studied. The effective 
Hamiltonian is given by
Eq. (\ref{model}). Both phases, the BEC and the MI phase, were found within the same mean-field 
approach to our model.

In the BEC phase there is a long-range order in the phase fluctuations (i.e. phase coherence) and
quasiparticles have a linear excitation spectrum for small momenta $q$. This reflects the existence 
of a Goldstone mode due to the breaking of a global $U(1)$ gauge symmetry. Approaching the MI
phase the system looses its phase coherence at the transition point and the excitation 
spectrum is characterised by the gap opening.

In previous calculations, performed on the Bose-Hubbard model, each phase requires its
own specific mean-field approach \cite{stoof,liu} or a single one close to the phase boundary 
\cite{huber}. 
In the MI phase our expression for the excitation spectrum agrees with the branch of the 
excitation spectrum, which 
corresponds to creation of holes    
in the first Mott lobe of the Bose-Hubbard model, taking the limit of a large interaction. 
However, the second branch, which corresponds to formation of doubly occupied sites, does not exist in 
our model since we can create only holes in
the singly occupied lattice
to excite our system and not additional particles due to the hard-core condition. This is possible
in the grand-canonical ensemble, where only the average number is fixed but the number of particles
fluctuates.  

Within a Bogoliubov 
approximation to the Bose-Hubbard model
the quasiparticle spectrum in the BEC phase was found as
\cite{stoof,liu}
\[
\epsilon_{q}=\sqrt{J^2 g_{q}^2+2U n_{0} J g_{q}},
\]    
where $U$ is the interaction parameter and $n_{0}$ is the condensate density. In contrast to this expression,
we found for the spectrum the expression in Eq. (\ref{spectrum}). These expressions do not agree in the
limit $U\to\infty$. Thus our hard-core Bose gas cannot be described within the Bogoliubov 
approximation to 
the Bose-Hubbard model by simply sending $U$ to infinity. On the other hand, our results are in good 
agreement with a variational Schwinger-boson mean-field approach to the Bose-Hubbard model, which 
describe the phases near the phase transition, by sending 
$U$ to infinity \cite{huber}.

By changing the particle
density from low to high values we pass from the weakly to strongly interacting regime.
Within our approximation we were able to calculate an effect of quantum and thermal fluctuations
on the condensate. The thermal fluctuations have strong effect: They destroy the phase boundaries in a
direction of its vanishing. For fixed tunneling rate $J$ and chemical potential $\mu$ one can
determine a critical temperature at which the condensate vanishes. The quantum fluctuations do not
change phase boundaries.

\subsection{Critical Exponents}

Our expansion of the action $A_{\rm eff}$ near the phase
transition between the BEC and the Mott insulator Eq.(\ref{Landau}) is similar to the Ginzburg-Landau 
theory of (thermal) phase transitions if we use the relation
\begin{equation}
\frac{T-T_{c}}{T_{c}}\rightarrow
~\frac{\mu-J}{J}
\end{equation}
and $\mu_{c}=J$, i.e. the chemical potential $\mu$ plays
the role of the temperature $T$ in our zero-temperature phase diagram.
Within our approximation we find the mean-field expressions for three
critical exponents: The critical exponent of the correlation
length $\xi$ of density fluctuations is $\nu=1/2$, of the compressibility $\chi$ it is
$\gamma=1$ and of the order parameter $\phi$ it is $\beta=1/2$. These critical 
exponents characterize the divergence of
the corresponding quantities at the phase transition:
$$
 \xi\sim (\mu-J)^{-\nu} \; , \quad
 \chi\sim (\mu-J)^{-\gamma} \; , \quad
 \phi\sim (\mu-J)^\beta \; .
$$

\subsection{Static Structure Factor}

One possibility to detect the phase transition between the BEC and the MI in an optical
lattice is to probe the excitation spectrum \cite{exper1, Esslinger}.
In the superfluid phase a broad continuum of excitations is observed but
in the MI phase a more structured spectrum is measured, indicating the existance of the gap. The 
static
structure factor vanishes in the Mott phase, as can be detected in experiments
when studying phase transitions in an optical lattice.

We also notice that the particle density cannot serve as an order parameter for the BEC-MI transition 
and the
scaling law is not applicable for density-density correlations, which can be determined as a Fourier
transform of the static structure factor and is given in Appendix B. However, we can assume the
validity of the scaling law for
correlations of the order parameter, namely
\begin{equation}
\langle \delta\phi_{r}^{\ast}\delta\phi_{r^{\prime}} \rangle.
\end{equation}
Knowing the Green's function (see Appendix A) we obtain the
following expression:
\begin{equation}
\langle
\delta\phi_{p}\delta\phi_{-p^{\prime}}\rangle=\delta_{p,p^{\prime}}
G_{p}\sim \delta_{p,p^{\prime}}\frac{1}{\epsilon_{q}}.
\end{equation}
Then for large distances $r$ and for $J=\mu$ (i.e. at the phase transition) we find
the power law
\begin{equation}
C_{0,r}\sim\int d^d r \langle \delta\phi_{p}\delta\phi_{-p}\rangle
e^{iqr} \sim \int d^{d}q\frac{const.}{q^{2}}e^{iqr}\sim
\frac{1}{r^{d-2}} \; ,
\end{equation}
which gives the anomalous exponent $\eta=0$ as in a mean field Landau-Ginzburg theory.

\section{Conclusions}

We have used a paired-fermion model to describe strongly interacting bosons in an optical lattice
with hard-core interaction. On the level of a mean-field theory we calculate the phase diagram, 
which includes the BEC and the MI.
Including Gaussian fluctuations, we have found that the dispersion of quasiparticles
is gapless in the BEC phase but has a gap in the MI phase:
\[
\epsilon^{\rm BEC}_{q}=\sqrt{g_{q}(J^2-\mu^2)+g_{q}^2\mu^2},
\]
\[
\epsilon^{\rm MI}_{q}=\mu-J+J g_{q} .
\]
$g_{q}$ is the dispersion of the bosons on the lattice, defined in Eq. (\ref{g}). 

We have calculated the total density, the condensate density,
and the static structure factor. We have shown that the quantum fluctuations as well as thermal fluctuations  lead to a depletion of the 
condensate, but the former do not change the
critical points.
The static structure factor contains information about the quasiparticle 
excitations and in the BEC phase for small $q$ it is
\[
S(q)\sim\frac{(J^2-\mu^2)}{J^2}\frac{J g_{q}}{\epsilon_{q}}\coth\frac{\beta\epsilon_{q}}{2} .
\]
It vanishes in the MI phase.
For the critical behavior of the compressibiliy, the
density correlations and the order parameter at the phase transition between the BEC and
the MI phase we found typical mean-field results.

\appendix
\section*{Appendix A: Green's function}
In this Appendix we write out the expression for the Green's function
in both cases $|\phi|=0$ and $|\phi|\ne 0$. We denote $p=\{q,\omega\}$.
\subsection*{Case: $|\phi|=0$}
Deviation of the effective  action due to fluctuations is
\begin{equation}
\delta A_{\rm eff}=\sum_{p}(\begin{array}{cc} \delta \phi_{p} \ \ \delta\chi_{p}  
\end{array})\overbrace{\left(\begin{array}{cc} v^{-1}_{p}-D(p) & i D(p) \\  i D(p) & 
\frac{1}{2J}+D(p) 
\end{array}\right)}^{G^{-1}}\left(\begin{array}{c}\delta\bar{\phi}_{p} \\ \delta\bar{\chi}_{p}   
\end{array}\right),
\end{equation}
where
\[
D(p)= \frac{1}{|\mu|-i\omega}, \ \ v^{-1}_p=\frac{1}{J(3-g_q)}.
\]
The determinant of the Green's function reads
\begin{equation}
\det
G^{-1}=\frac{v^{-1}_{p}}{2J}-D(p)\left(\frac{1}{2J}-v^{-1}_{p}
\right).
\end{equation}
\subsection*{Case: $|\phi|\ne 0$}
Deviation of the effective action due to fluctuations is
\begin{equation}
\delta A_{\rm eff}
=\sum_{p>0}(\begin{array}{cccc} \delta \phi_{p}, \delta\chi_{p}, \delta\bar{\phi}_{-p}, \delta\bar{\chi}_{-p}  
\end{array})G^{-1}\left(\begin{array}{c}\delta\bar{\phi}_{p} \\ \delta\bar{\chi}_{p}\\ \delta\phi_{-p}\\ \delta\chi_{-p}  
\end{array}\right)
\end{equation}
with the Green's function
\begin{equation}
G^{-1}=\left(\begin{array}{cccc} v^{-1}_p-D(p) & i D(p) & -a & i a \\ i D(p) & 
\frac{1}{2J}+D(p) & i a & a \\
 -a & i a &  v^{-1}_p-D(-p) & i D(-p)  \\ i a & a &  i D(-p) & \frac{1}{2J}+D(-p)  
\end{array}\right),
\end{equation}
where
\[
D(p)=\frac{1}{2}\cdot\frac{\mu^{2}+J^{2}+2i\mu\omega}{J(J^{2}+\omega^{2})},
\]
\[
D(-p)=\frac{1}{2}\cdot\frac{\mu^{2}+J^{2}-2i\mu\omega}{J(J^{2}+\omega^{2})},
\]
\[
a=-\frac{1}{2}\cdot\frac{|\Phi|^{2}/9}{J(J^{2}+\omega^{2})}.
\]
The determinant of the Green's function is
\begin{equation}
\det G^{-1}=\frac{1}{[2J^{2}(3-g_q)]^{2}(J^{2}+\omega^{2})}
\cdot [\omega^{2}+(J^{2}-\mu^{2})g_{q}+\mu^{2}g_{q}^{2}].
\end{equation}

\section*{Appendix B: Correlations}
The decay of the density-density correlation function can be investigated as
the inverse Fourier transform of the static structure factor
\[
C_{r,0}\sim \frac{(J^2-\mu^2)}{J^2}\int d\vec{q} \frac{Jg_{q}}{\epsilon_{q}}e^{i\vec{q}\vec{r}}
\]
For large values of $r$ the main contribution to the integral is for small values of $q$, namely
\begin{equation}
\lim_{r\rightarrow \infty}C_{r,0}\sim \int dq d\Omega q^{d}e^{i\vec{q}\vec{r}}\sim \frac{1}{r^{d+1}}\int dq 
q^{d-1}\sin(q)\sim \frac{1}{r^{d+1}}.
\end{equation} 

\bibliography{reference} 

\begin{figure}[h]
\begin{center}
\includegraphics[width=12cm]{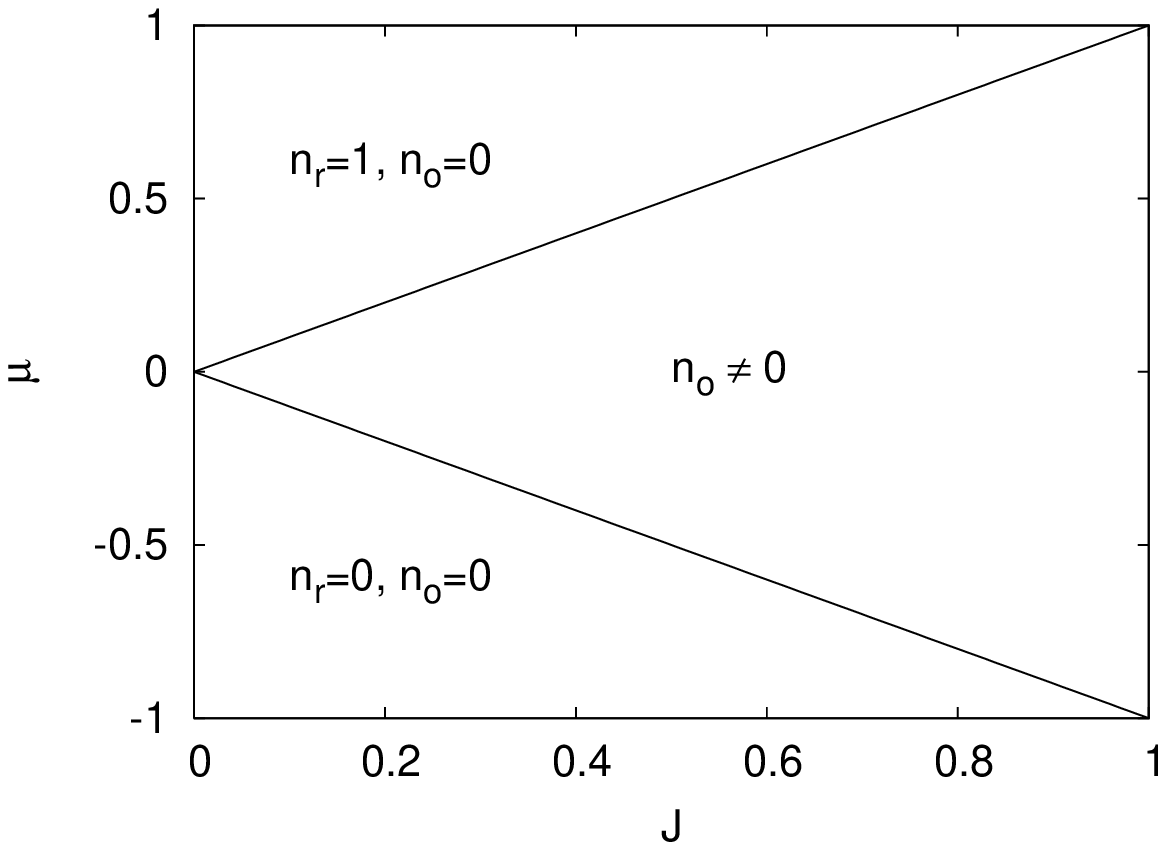}
\end{center}
\caption{Phase diagram for $k_B T=0$. $\mu$ and $J$ are in arbitrary energy units.}
\label{phasediag}
\end{figure}

\begin{figure}[h]
\begin{center}
\includegraphics[width=12cm]{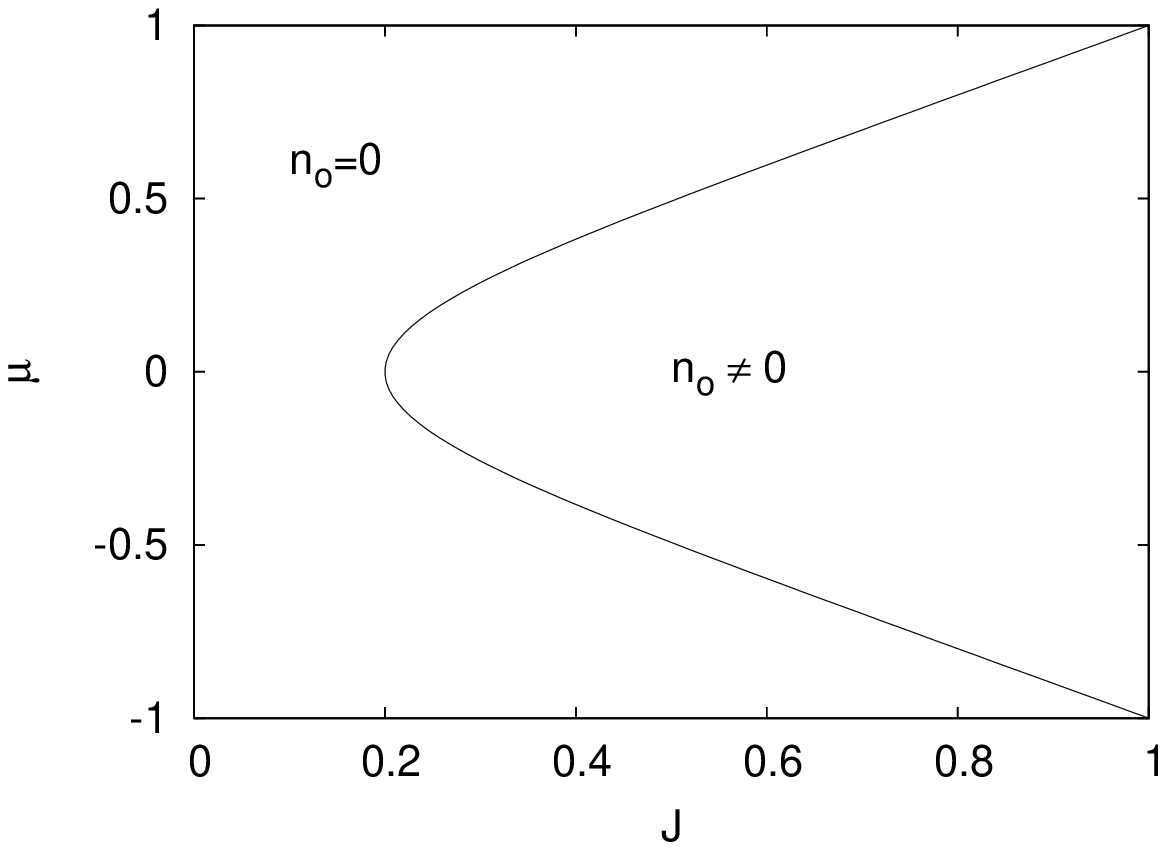}
\end{center}
\caption{Phase diagram for $k_B T=0.05 J$. $\mu$ and $J$  are in arbitrary energy units.}
\end{figure}

\begin{figure}[h]
\begin{center}
\includegraphics[width=12cm]{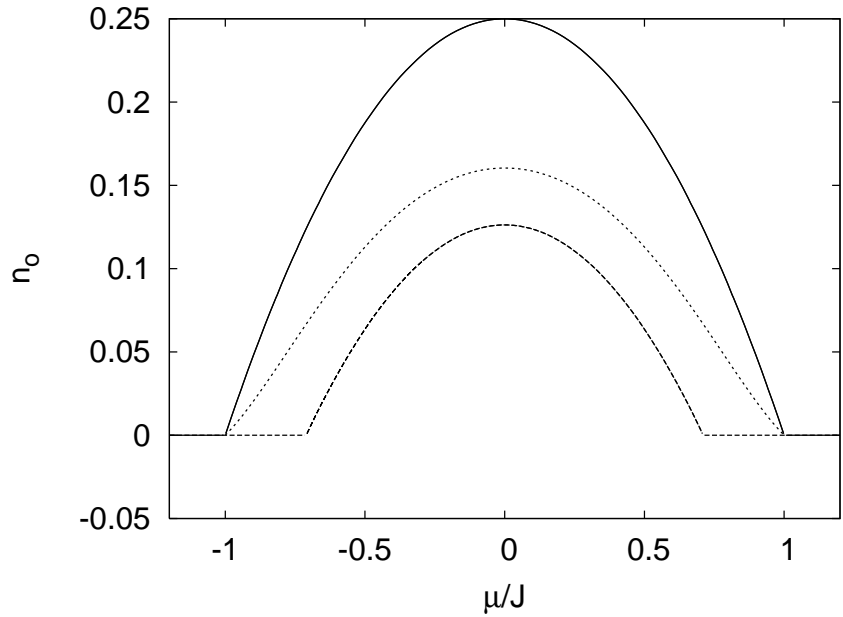}
\end{center}
\caption{Condensed density. The solid, dotted and dashed lines show the mean-field result at $k_B T=0$,  
the influence of the quantum 
fluctuations at $k_B T=0$ to the mean-field result and the mean-field result at $k_B T=0.2 J$, respectively. 
$\mu$ and $J$ are in arbitrary energy units.}
\label{fluct}
\end{figure}

\begin{figure}[h]
\begin{center}
\includegraphics[width=12cm]{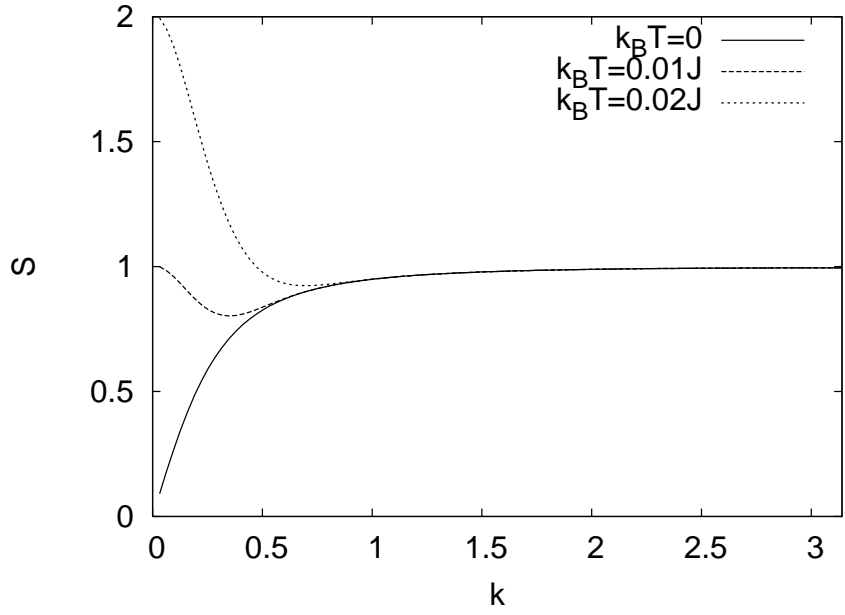}
\end{center}
\caption{Static structure factor in the BEC phase in the dilute regime, $\mu = -0.99 J$. $J$ is in arbitrary 
energy units. The unit of length is the lattice constant, so $k$ is dimensionless.}
\label{Sdilute}
\end{figure}

\end{document}